\documentclass[aps,pre,notitlepage,amsmath,amssymb,amsfonts,floatfix,superscriptaddress]{revtex4-1}

\usepackage{graphicx}
\usepackage{color} 
\usepackage[colorlinks=true, citecolor=blue,urlcolor=blue,linkcolor=blue,filecolor=black]{hyperref}
\usepackage{multirow}
\usepackage[T1]{fontenc}
\usepackage{dsfont}
\usepackage{hyperref} 
\usepackage{cleveref} 
\usepackage{floatrow}  
\usepackage{natbib}
\usepackage{epstopdf}
\usepackage{ifpdf}
\usepackage{epsfig}
\usepackage[normalem]{ulem}
\usepackage[usenames,dvipsnames]{xcolor}
\usepackage{makecell}
\usepackage{upgreek}
\usepackage{fixltx2e}

\usepackage{empheq}

\begin{document}

\title{Drag force on cylindrical intruders in granular media: Experimental study of lateral vs axial intrusion and high grain-polydispersity effects}

\author{Salar Abbasi Aghda}\thanks{salar.abbasi@ipm.ir}
\affiliation{School of Nano Science, Institute for Research in Fundamental Sciences (IPM), P.O. Box 19395-5531, Tehran, Iran}
\affiliation{Condensed Matter National Laboratory, Institute for Research in Fundamental Sciences (IPM), P.O. Box 19395-5531, Tehran, Iran}
\author{Ali Naji}\thanks{a.naji@ipm.ir (corresponding author)}
\affiliation{School of Nano Science, Institute for Research in Fundamental Sciences (IPM), P.O. Box 19395-5531, Tehran, Iran}

\begin{abstract}
We report on experimentally measured drag force experienced by a solid cylinder penetrating a granular bed of glass beads including, specifically, a highly polydisperse sample with grain sizes in the range 1-100~$\upmu$m. We consider both cases of lateral and axial intrusion in which the long axis of the cylinder is held horizontally and vertically, respectively. We explore different regimes of behavior for the drag force as a function of the penetration depth. In lateral intrusion, where the motion is effectively quasi-two-dimensional, we establish quantitative comparisons with existing theoretical predictions across the whole range of depths. In axial intrusion, we observe peculiar undulations in the force-depth profiles in a wide range of intrusion speeds and for different cylinder diameters. We argue that these undulations reflect general shear failure in the highly polydisperse sample and can be understood based on the ultimate bearing capacity theory on a semiquantitative level. 
\end{abstract}

\maketitle

%%%%%%%%%%%%%%%%%%%%%%%%%%%%%%%%%%%%
%%%%%%%%%%%%%%%%%%%%%%%%%%%%%%%%%%%%
\section{Introduction}

Granular materials are collections of macroscopic solid particles or `grains' (such as sand, glass beads, etc.) in substantial interparticle contacts in a way that no individual grain can move independently of its neighbors \cite{durian,RevModPhys.68.1259,herminghaus2005dynamics,antony2004granular,herrmann2002granular,ness2020absorbing}. The gaps between the grains is typically filled by air or other fluids. Unlike the traditional continuum matter (such as Newtonian fluids), the rheology of granular materials is complicated by their internal inhomogeneity which implies comparatively different force transmission  mechanisms \cite{chen2010rheology,PhysRevLett.82.205,PhysRevE.53.4673,Majmudar2005ContactFM,RevModPhys.71.S374}. Externally applied forces are thus transmitted through force chains formed within the medium by random contacts of neighboring particles. A particular situation where such force transmissions are important is the motion of a macroscopic object (intruder) in a granular bed which is a common motif in geophysics, fluid dynamics and robotics. Examples include asteroid impact events \cite{PhysRevLett.91.104301,Nelson2008,PhysRevLett.109.238302}, industrial drilling and mixing or segregation  of minerals \cite{ELLENBERGER2006168}, silo monitoring \cite{PhysRevE.74.011307,BRIDGWATER2012397,PhysRevLett.106.218001,doi:10.1146/annurev.fluid.32.1.55} and subsurface animal locomotion  \cite{maladen2009undulatory}. 

While the force experienced by solid objects moving in viscous fluids at low Reynolds numbers is well established  \cite{happel2012low}, the effect in granular systems is much less explored and has received growing attention only in the recent past \cite{PhysRevLett.82.205,stone2004getting,PhysRevE.70.041301,Hill_2005,peng2009depth,B926180J,PhysRevLett.106.028001,kolb2013rigid,brzinski2013depth,PhysRevE.87.052208,kang2018archimedes,C8SM02480D}. In low-speed granular impact processes, the total drag (or resistance) force on an intruder at a given depth is found to be a combination of two different contributions; a frictional drag that increases linearly with the depth of intrusion and an inertial drag that increases quadratically with the speed of intrusion \cite{Katsuragi2007UnifiedFL}.

For quasistatic penetration (low velocities), the frictional drag is responsible for the total drag which thus behaves independently of penetration velocity.  This is because the contribution from inertial drag turns out to be zero for penetration velocities lower than a critical value $V_\text{c}=\sqrt{2gd}$. The latter corresponds to the velocity of a grain falling down a distance equal to its diameter $d$ under the gravitational force \cite{PhysRevLett.82.205,clark2013granular,2016NatPh..12..278A}. Recent impact studies have however indicated a more complex picture than predicted by the additive frictional-inertial drag model. For instance, when impact velocities are above $V_\text{c}$, the velocity-squared dependence of drag force is indeed observed (as pronounced peaks in force-depth curves) but only for a short period of time after the impact. Afterwards, the drag  reduces to a purely  depth-dependent frictional  force \cite{PhysRevLett.126.218001,roth}. On the other hand, it has been shown that the additive force model is accurate only near a critical packing fraction \cite{PhysRevE.82.010301}. 

Frictional drag has also been studied in relation to criteria on infinite falling motion of spherical objects in long columns of polydisperse beads \cite{PhysRevLett.106.218001}. In this case, the existence of a terminal velocity  has  been established for intruder mass above a critical value.  Effects from other factors such as confinement due to a container have also been investigated. The drag force is modified upon close approach to the bottom of container \cite{Nelson2008,seguin2008influence}. 

In this work, we experimentally investigate the drag force on a solid cylindrical intruder with the purpose of elucidating some of the less explored aspects of the problem. In our experiments, we  use a  custom-made automatic device consisting of a motorized linear stage that can vertically push the  intruder  at constant speed into a container filled with glass beads. We  consider both lateral and axial intrusions where the cylinder is held horizontally and vertically as it penetrates the granular bed, respectively. The former case can be viewed as a quasi-two-dimensional setting. In this case, we limit our study to the quasistatic regime ($V<V_\text{c}$) and arrange the experimental setting according to Ref. \cite{PhysRevLett.107.048001} where the frictional drag is studied in the so-called hydrostatic-like regime (i.e., relatively small depths with  negligible container effects). To the best of our knowledge,  quantitative comparisons between experiments and existing theoretical models \cite{brzinski2013depth}  in this quasi-2D setting are still missing. In our experiments, not only we establish such a comparison between our data and  the theoretical predictions in the hydrostatic-like regime, we also go further by exploring the whole range of intrusion depths from the surface down to the vicinity of the bottom wall. 

In axial intrusion (vertically oriented cylinder), we focus on shallow penetration depths  and study two relatively monodisperse granular samples as well as a highly polydisperse sample. The former samples are primarily included for validation purposes (successful comparisons between theory and experiments in such cases have previously been reported  \cite{stone2004getting,PhysRevE.70.041301,brzinski2013depth,kang2018archimedes,PhysRevLett.126.218001}). The highly polydisperse sample presents an intriguing situation that has not been scrutinized before. Here, we report  on the dependence  of drag force on the intrusion velocity and cylinder diameter. The force-depth profiles display pronounced undulations with the intrusion depth. We argue that these undulations reflect the bearing capacity of granular sample in the highly polydisperse case.

%%%%%%%%%%%%%%%%%%%%%%%%%%%%%%%%%%%%%%%
%%%%%%%%%%%%%%%%%%%%%%%%%%%%%%%%%%%%%%%
\section{Experimental Setup}

%%%%%%%%%%%%%%%%%%%%%%%%%%%%%%%%%%%%
\subsection{Materials and methods}

\begin{figure}[t!]
    \centering
    \includegraphics[width=0.5\textwidth]{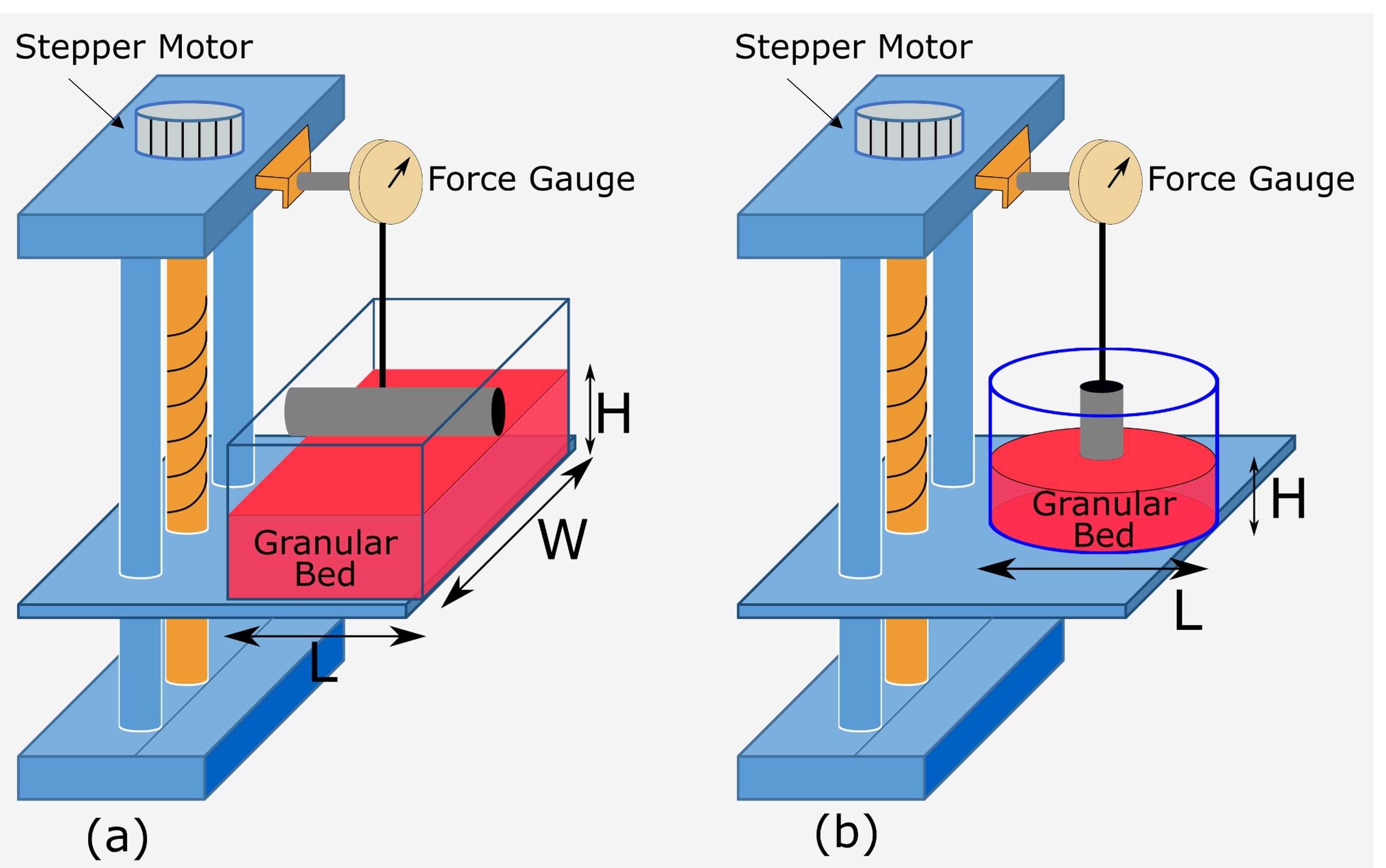}
    \caption{Schematic view of the experimental setup used for measuring the drag force imparted on a solid cylindrical intruder penetrating into a rectangular (a) or cylindrical (b) container filled with glass beads. The long axis of intruder is kept  fixed during the penetration. 
    }
    \label{fig:schematic}
\end{figure}

 To study the importance of different factors contributing to the drag force experienced by an intruder moving in a granular bed, an automatic setup was designed and built as schematically pictured in Fig. \ref{fig:schematic}. The custom-made apparatus consists of a motorized linear translation stage, a force gauge, a solid cylinder made of  aluminum as the intruder (shown in gray), and a granular bed of glass beads (shown in red) in a glass container. The intruder is connected to the force  gauge by a vertical thin rod and its penetration into the granular bed is rendered by the upward motion of the container.  

%%%%%%%%%%%%
%%%%%%%%%%%%

The vertical  translation  is enabled  by the motorized linear stage consisting of three stainless steel bars. The two side bars (in light blue) are fixed and held together with top and bottom end-clamps. These make a rail of length 50~cm that maintains and stabilizes movement of the granular box holder.  The middle bar (in orange) is a lead screw of pitch 1~cm, converting the rotary motion of the stepper motor into translational motion of the linear stage. The positioning and speed control of the container are achieved using a stepper motor (Leadshine 86HS45 2-Phase NEMA 34) installed at the end of the linear motion rail. The stepper motor is connected to the lead screw by an aluminum star coupling and is capable of producing torques up to 4.5~N$\cdot$m. The stepper motor and the designed lead screw enable linear speeds $V$ up to 50~$\text{mm}/\text{s}$. The  motion parameters (speed, acceleration, pauses, penetration depth, etc.) are tuned using Leadshine MA860H Microstepping Driver that sends programmable control pulses to the stepper motor.  We use an IMADA ZTA 100 N force gauge with both push and pull measuring capability. The force experienced by the intruder is  thus monitored while the speed of penetration is tuned by the designed control system. Upon connection to a computer or a flash memory, the force gauge is capable of processing and storing data with a sampling rate of 2~kHz at a precision of $\pm$0.1~N.

%%%%%%%%%%%%
\begin{table*}[t!]
\centering
\caption{Parameter values for different granular samples used in our experiments.}
\label{tab:my-table}
\resizebox{0.95\textwidth}{!}{
\begin{tabular}{>{\centering}m{0.15\textwidth}>{\centering}m{0.15\textwidth}>{\centering}m{0.25\textwidth}>{\centering}m{0.15\textwidth}c}
\hline 
Sample & \thead{Size \\ $d\,$(mm)} & \thead{Apparent density \\ $\rho\, (\text{g/cm\textsuperscript{3}} )(\pm 0.01)$}   & \thead{Packing fraction \\ $(\pm 0.005)$}  & \thead{Angle of repose \\ $\theta\,$($\pm 2^{\circ}$)} \\
\hline
Glass beads 1 & 2         & 1.52    & 0.608  & 20    \\
Glass beads 2 & 0.4-0.6   & 1.40    & 0.560  & 22    \\
Glass beads 3 & 0.001-0.1   & 1.30    & 0.520  & 27    \\
\hline
\end{tabular}
}
\end{table*}
%%%%%%%%%%%%

The experiments were conducted while the ambient temperature and humidity were kept at  $(30\pm 1)^{\circ}$C and $(29\pm 4)$\%, respectively. Three different samples of glass beads were used in the experiments; see Table \ref{tab:my-table}. These include two relatively monodisperse samples (1, 2) with bead sizes (diameters) $d=2$~mm and $d=400-600$~$\upmu$m and a highly polydispere sample (3) with bead sizes in the range  1-100~$\upmu$m. The material density of the grains in the  three samples is 2.5  g/cm\textsuperscript{3} and the apparent sample density $\rho$ (defined as material grain density multiplied by the packing fraction) for each case  is given in the table.   In order to maintain the same packing fraction in all experiments, equal mass of glass beads was used in each run. The granular bed was prepared by slowly pouring the beads into the container and by shaking it in such a way that the upper surface of the bed is always established at the same prescribed height $H$ from the bottom of the container. After each run of the force measurements, the container was fully emptied  and refilled again to enable randomized particle contacts in the sample. As seen in the table, the measured packing fractions and apparent densities  for the three samples are roughly the same. The angle of repose $\theta$ for each sample is also provided in the table. This angle is different for the three samples and is measured by slowly pouring the grains from a funnel onto an isolated flat surface and averaging the  steepest angle of descent with the horizontal plane over 5 runs.

%%%%%%%%%%%%%%%%%%%%%%%%%%%%%%%%%%%%
\subsection{Designs for lateral vs axial intrusion}
\label{subsec:designs}

We conducted two different sets of experiments with  the cylindrical intruder being kept at fixed horizontal orientation (lateral intrusion, Fig. \ref{fig:schematic}a) and at fixed vertical orientation  (axial intrusion, Fig. \ref{fig:schematic}b). 

In lateral intrusion, we use a rectangular  glass container of inner side lengths $L\times W =50~\text{mm}\times  100~\text{mm}$ filled  with glass beads of size $d=2$~mm (sample 1 in Table \ref{tab:my-table}) up to the height $H= 180$~mm. The cylinder is of length 49~mm and diameter $D=20$~mm (the millimeter-sized gaps between the cylinder bases and the container walls prevent wall friction and grain trapping). The penetration speed is kept fixed at $V=1$~mm/s. In this case, the intrusion proceeds from the top surface of the granular bed at $Z=0$ to close proximity of its bottom at $Z= H$, where $Z$ is the distance between the bottommost point of the cylinder and the bed surface.

In axial intrusion, we use a cylindrical glass container of diameter $L=140$~mm. In this case, we use three different cylindrical intruders with equal  length 50~mm and diameters $D=10, 20$ and 30~mm.  For symmetry reasons, the cylindrical intruder is held coaxially with the container. In this case, we use all three samples from Table \ref{tab:my-table} and similarly fill the container  up to the height $H= 180$~mm. 
Here, we focus on the regime of relatively shallow penetration depths  to avoid the bottom wall effects and explore the role of penetration speed by increasing it from  $V=1$ up to 40~mm/s. The experiments corresponding to each reported data in this work are independently repeated up to 9 times  to establish reproducibility and estimate error bars.  

%%%%%%%%%%%%%%%%%%%%%%%%%%%%%%%%%%%%
%%%%%%%%%%%%%%%%%%%%%%%%%%%%%%%%%%%%
\section{Results and discussions}

%%%%%%%%%%%%
\begin{figure}[t!]
    %\centering
    \includegraphics[width=0.5\textwidth]{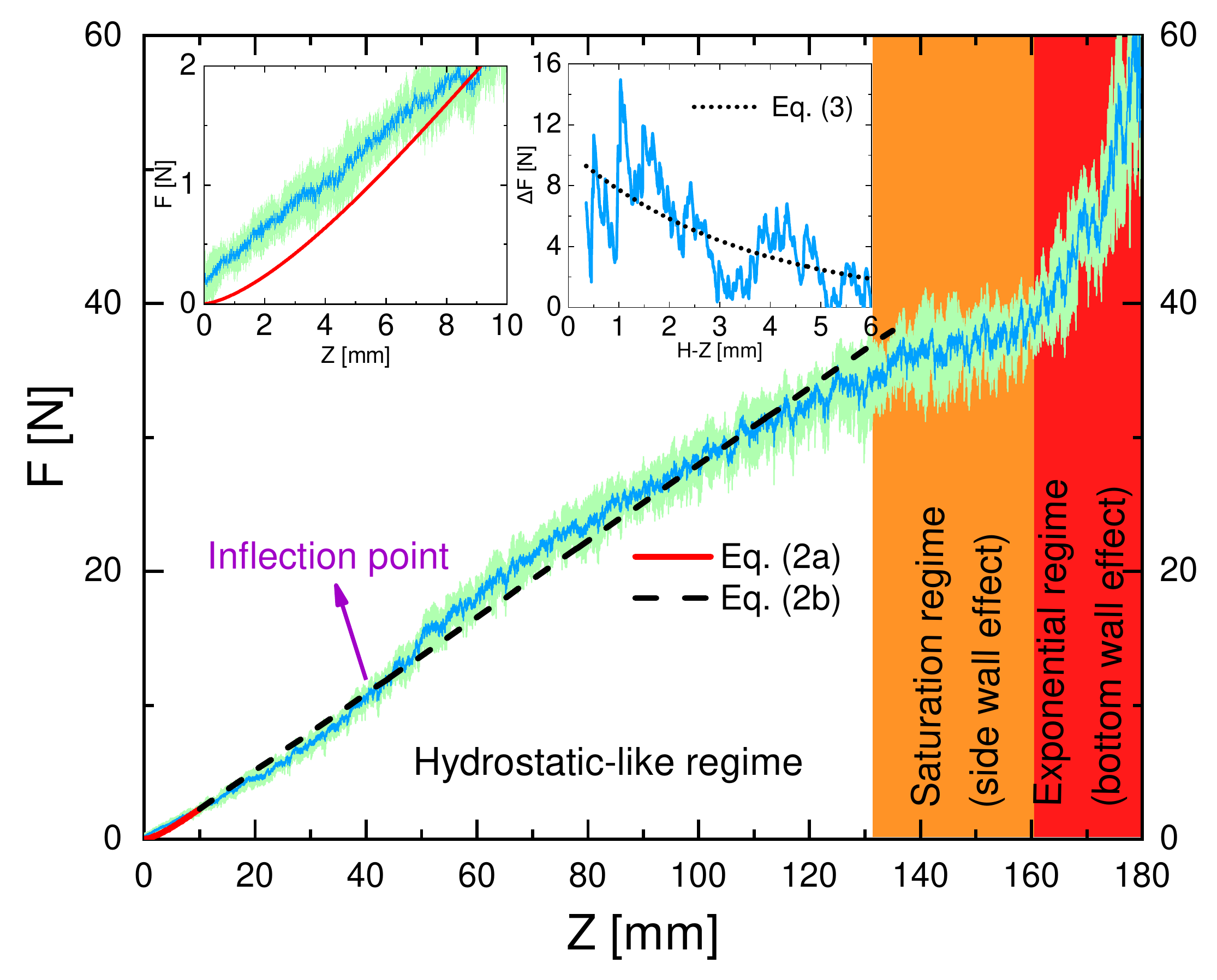}
    \caption{Measured drag force as a function of intrusion depth for a horizontally oriented cylinder  of diameter $D=20$~mm penetrating a bed of glass beads (sample 1) at constant speed $V=1$~mm/s  (quasi-2D setting, Fig. \ref{fig:schematic}a).  The blue curve shows the mean force with shaded error region shown in green. The solid red and dashed back lines show  the best fits to Eq. \eqref{eq1} and  Eq. \eqref{eq2}, respectively. Left inset shows magnified view of small-depth region ($0<Z<R$) and the right inset shows the large-depth behavior near the bottom wall along with the respective theoretical fit from Eq. \eqref{exp}.}
    \label{section-plot}
\end{figure}
%%%%%%%%%%%%

%%%%%%%%%%%%%%%%%%%%%%%%%%%%%%%%%%%%
\subsection{Drag force on laterally intruding cylinder}
 
Figure \ref{section-plot} depicts the measured drag force, $F$, as a function of the penetration depth, $Z$, on a horizontally orientated cylinder with parameter values described above. The blue curve shows the mean drag force with shaded error region being shown in green. The force-depth profile indicates three different regimes of behavior shown with white, orange and red backgrounds. 

The first regime (white) reflects the hydrostatic-like behavior where the force increases almost  linearly with the penetration depth. The linearity is visualized by the dashed back line in Fig. \ref{section-plot}, main set, which is the best fit to the data (see Eq. \eqref{eq:all} below).  
 
In this regime, Brzinski  et. al  \cite{brzinski2013depth} showed that local friction forces exerted by the surrounding grains on the surface of an intruding object produce a local frictional drag  as
\begin{equation}
    \mathrm{d}F=\alpha\mu(\rho g Z)\,\mathrm{d}A, 
    \label{friction}
\end{equation}
where $\mu=\tan \theta$ is the relevant friction coefficient, $\mathrm{d}A$ is the area element for the surface of the intruder, $\rho gZ$ is the hydrostatic pressure at depth $Z$ due to the gravitational acceleration $g$, $\rho$ is the apparent density of the granular bed as noted before, and $\alpha$ is a prefactor which appears to be independent of intruder shape \cite{stone2004getting,PhysRevE.70.041301,brzinski2013depth}. For ordinary Coulomb friction, this prefactor is expected to be of the order of unity. The experiments however indicate a value larger by an order of magnitude \cite{stone2004getting,PhysRevE.70.041301,brzinski2013depth}. This is because the gravity-loaded contacts lead to a greater effective area due to the force chains that spread out from the intruder surface    (this is further confirmed by object withdrawal observations where the force chains are infrequent and $\alpha\sim 1$ \cite{Hill_2005}). 

For the case under study in this section, Eq. \eqref{friction} yields $F=\alpha\mu (\rho gL)R B$ where $L$ and $R=D/2$ are cylinder length and radius, respectively, and $B$ is obtained as \cite{brzinski2013depth}

%\scriptsize{}
\begin{subequations}
\label{eq:all}
    \begin{empheq}[left={B=\empheqlbrace\,}]{align}
      & \sqrt{\left(2-\frac{Z}{R}\right)\frac{Z}{R}}\left(Z-R\right)+R\cos^{-1}\left(1-\frac{Z}{R}\right)  &: Z<R, \label{eq1}\\
      & 2Z-\left(2-\frac{\pi}{2}\right)R  &: Z>R. \label{eq2}
    \end{empheq}
\end{subequations}

\normalsize 
The solid red and the  dashed back lines in Fig. \ref{section-plot} (left inset and main set) show  the best fits to the data based on Eq. \eqref{eq1} and Eq. \eqref{eq2}  at depth intervals  $Z<R$ and $Z>R$, respectively. These fits are achieved for $\alpha= 55 \pm 5$; a value that agrees with the general range ($\alpha\sim 25-70$) reported previously. These reports include spherical, conical and axially intruding cylinders \cite{brzinski2013depth,Katsuragi2007UnifiedFL}. Our data thus largely confirm the experimental validity of  Eq. \eqref{eq:all}  in the quasi-2D case of a laterally intruding cylinder which was not addressed before. It is worth mentioning that the data for $Z<R$ (Fig. \ref{section-plot}, left inset) portray certain deviations  from the theoretical relation; i.e., while the data indicate a strictly linear dependence on $Z$, Eq. \eqref{eq1} predicts $F\sim Z^{3/2}$.  Such deviations are generally marginal and  appear also in the theory vs experiment comparisons made in other settings   \cite{brzinski2013depth}. The absolute deviations (up to around 0.5~N) are close to the precision of our force gauge (0.1~N). As such, they could be produced by measurement errors or other factors, including  surface phenomena that would require future clarification. Furthermore, the data in Fig. \ref{section-plot} reveal an inflection point (indicated by an arrow) at  $Z\simeq 40\,\text{mm}$. This is consistent with previous observations  on a spherical intruder  \cite{peng2009depth} where the inflection point was argued to arise at the depth of $Z\simeq 2R$. This depth is just  large enough to allow for the glass beads to reoccupy the top side of the intruder and partially compensate the opposing drag force. 

 The hydrostatic-like behavior is followed by a saturation regime at a depth of around $Z\simeq 135$~mm where the force levels off at a value of $F\simeq 37$~N. The saturation is a manifestation of the Janssen effect; i.e., the random force chains within the medium become relevant and partially transmit the apparent weight of the granular column above the intruder to the side walls of the container   \cite{seguin2008influence,sperl2006experiments,ZHAO2018149}. The Janssen effect is known to emerge at a depth roughly twice the ratio of the surface to the perimeter of the container \cite{sperl2006experiments}. In our case, this ratio is 120~mm and is thus in close agreement with the measured saturation  depth noted above. 
 
Beyond the saturation depth, experiments with an axially intruding disk \cite{stone2004getting,PhysRevE.70.041301} have revealed an exponentially increasing force within just a few layers of grains away from the bottom wall. This behavior can be expressed  as 
\begin{equation}
   \Delta F\propto \exp\left(\frac{H-Z}{\lambda}\right). 
\label{exp}
\end{equation}
Here, $\Delta F=F-F_\text{bulk}$ is the difference  between the measured drag force, $F$, and the corresponding force measured in a bulk system $F_\text{bulk}$ (i.e., in a sufficiently large container such that the bottom wall produces no discernible effects on the intruder). Also, $H$ is the granular column height and $\lambda$ is the characteristic distance over which the effects from the bottom wall come into play. To assess the applicability of this relation in our case, we find $F_\text{bulk}$ by extrapolating the hydrostatic-like line, Eq. \eqref{eq2}, to the the given depth $Z$ in the presumed exponential regime. Equation \eqref{exp} can be fitted to our data (dotted curve in the right inset of Fig. \ref{section-plot}) for a reasonable best-fit value of $\lambda=3.5$~mm which corresponds to $1.5$ layers of grains at the bottom. 

Although the foregoing regimes of hydrostatic-like behavior, force saturation and exponential growth have each  been reported separately in other intrusion settings in the past \cite{stone2004getting,brzinski2013depth}, our results in Fig. \ref{section-plot} establish them for the quasi-2D setting and concurrently through the entire range of intrusion depths.

%%%%%%%%%%%%
\begin{figure*}[t]
    \centering
    {\includegraphics[width=0.4\textwidth]{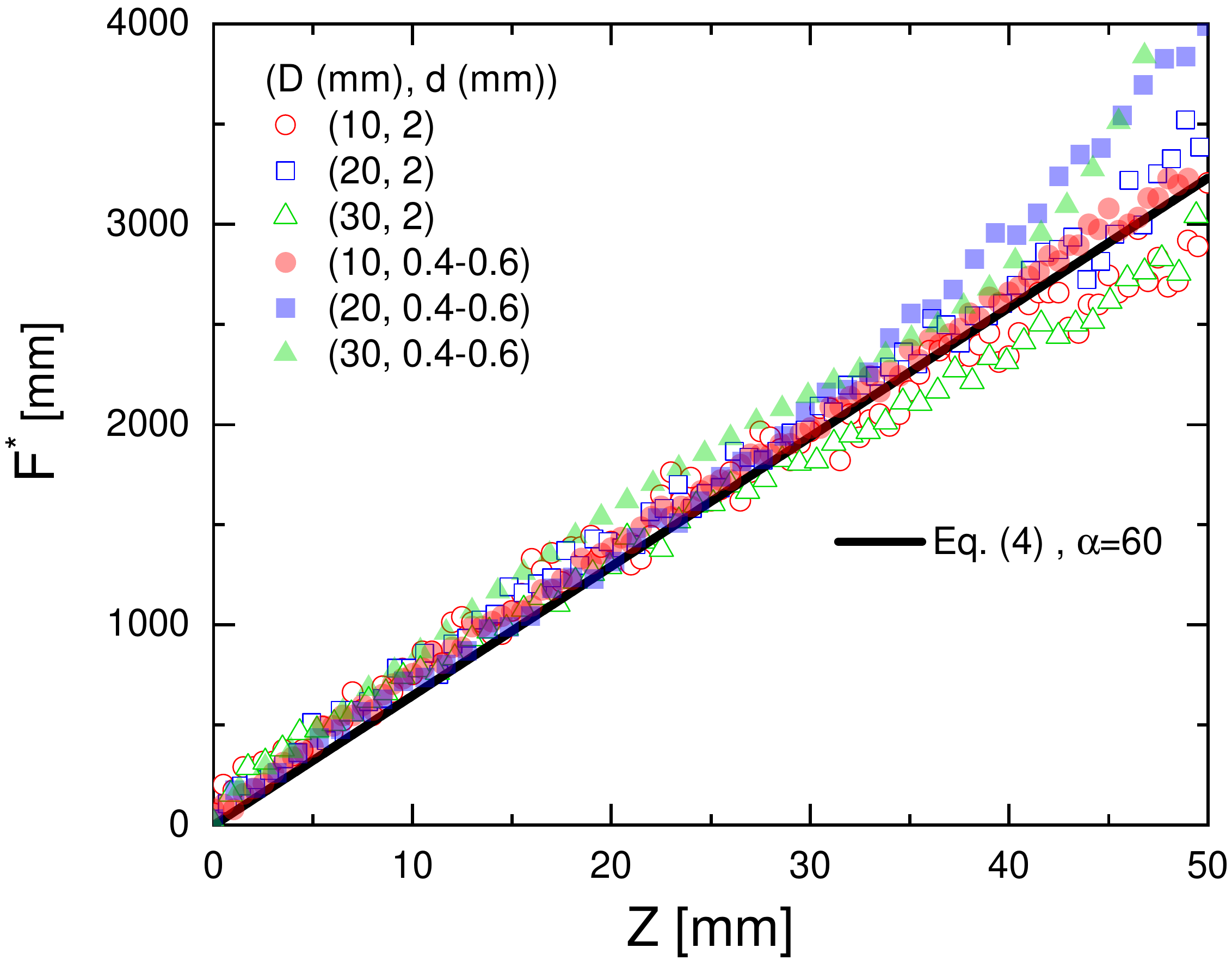}\label{PA2,400}} (a) \hskip10mm
    {\includegraphics[width=0.4\textwidth]{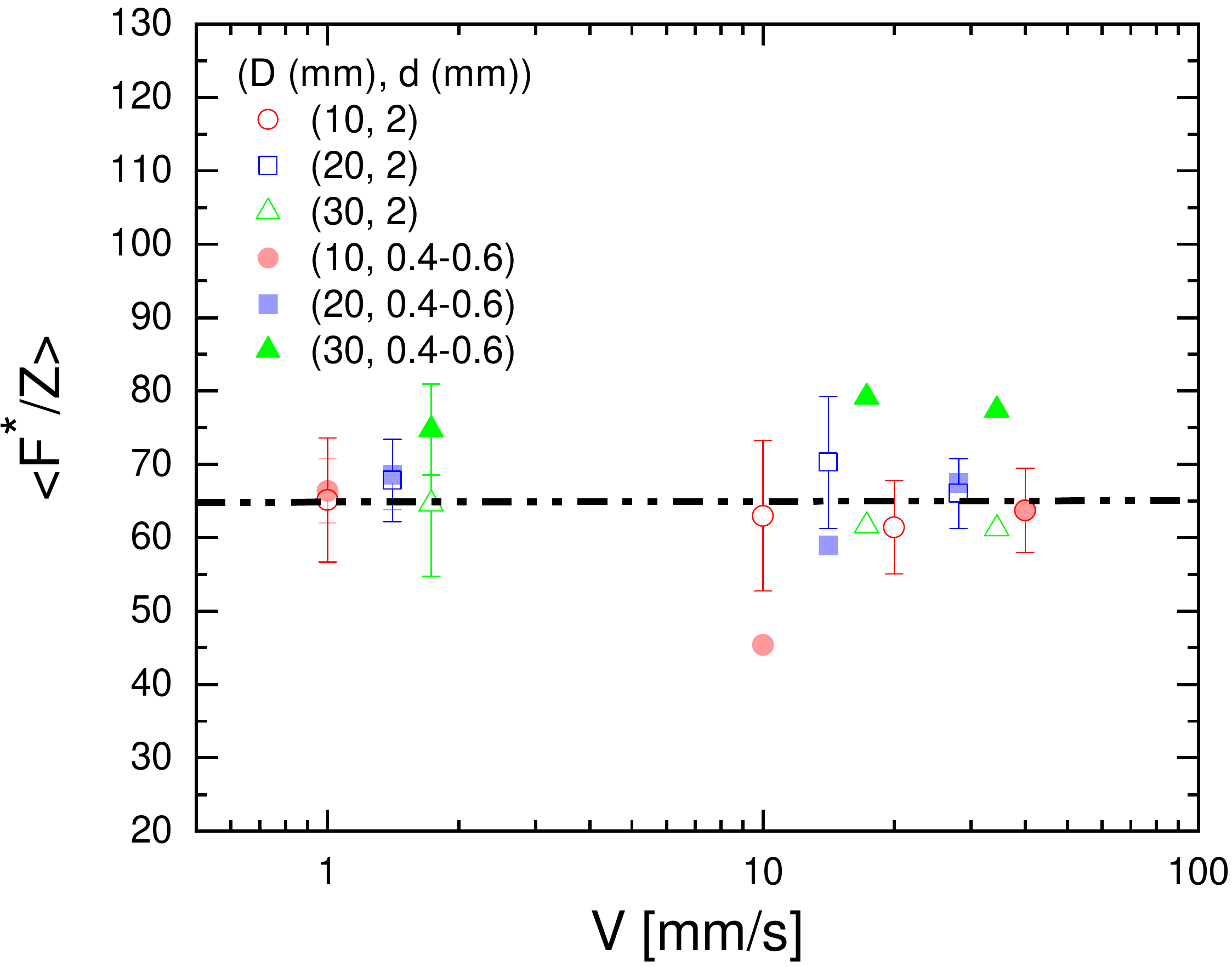}\label{PB2,400vel}} (b) 
    \caption{(a) Rescaled drag force $F^*=F/(\mu\rho gR^2)$ as a function of the intrusion depth for  cylinders of different diameters axially penetrating granular beds of samples 1 and 2 at constant speed $V=1$ mm/s (see also Fig. \ref{fig:schematic}b). Cylinder and grain diameters are indicated on the graph. The solid line shows the linear best fit to Eq. \eqref{VERTICAL}  with prefactor $\alpha=60$. (b) Semilogarithmic plot for $\langle F^*/Z\rangle$, obtained by averaging over individual runs of experiment, as a function of the intrusion speed.}
    \label{2mm,400}
\end{figure*}
%%%%%%%%%%%%

%%%%%%%%%%%%%%%%%%%%%%%%%%%%%%%%%%%%
\subsection{Drag force on axially intruding cylinder}

We now turn to the case of an axially intruding cylinder as described in  Section \ref{subsec:designs} (Fig. \ref{fig:schematic}b). In this case, the theoretical results in Ref.  \cite{brzinski2013depth} give the frictional drag as 
\begin{equation}
F=\alpha\mu\rho gR^2Z.
\label{VERTICAL}
\end{equation}
The experimental validity of this relation has been discussed in Refs. \cite{brzinski2013depth,kang2018archimedes}. While the former work assumes a constant $\alpha$ and $\mu=\tan \theta$, the latter argues that the linear dependence on $\tan\theta$ holds for repose angles around $22^{\circ}$ and that, beyond this value, the factor $\alpha\mu=K_\theta$ admits nonlinear corrections in terms of $\tan\theta$.  
Such corrections have been verified experimentally for samples with grain sizes in the range 100-1000~$\upmu$m \cite{kang2018archimedes}. The sample of interest in what follows (sample 3, Table \ref{tab:my-table}) is highly polydisperse and involves grain sizes varying over two orders of magnitude  below 100~$\upmu$m, not considered before.  However, to provide a wider perspective and connect more closely with previous works, we first consider  samples 1 and 2  (Table \ref{tab:my-table}) for which $\theta\simeq 22^{\circ}$ and the theoretical model of Ref. \cite{brzinski2013depth} is expected to hold. 

Figure \ref{PA2,400} shows the measured drag force on an axially intruding cylinder, being rescaled  as  $F^*=F/(\mu\rho gR^2)$, in the cases of samples 1 and 2 for three different cylinder diameters and constant intrusion speed $V=1$ mm/s. The data collapse well onto the line $F^*=\alpha Z$ in accordance with Eq. \eqref{VERTICAL} with best-fit value of $\alpha=60$ (solid black line). In Fig. \ref{PB2,400vel}, we repeat the experiments for different intrusion speeds, verifying that the ratio $\langle F^*/Z\rangle$ (averaged over each experimental run) remains nearly independent of the speed increased from $1$ to $40$ mm/s. Thus, even as part of our data is acquired at significantly larger speeds than those considered in Refs. \cite{brzinski2013depth,kang2018archimedes}, they still conform to the quasistatic scenario set forth by Ref. \cite{brzinski2013depth}. This is expected since the intrusion speed for samples 1 and 2 and for those in the said references remain consistently below the critical velocity $V_\text{c}=\sqrt{2gd}$. The foregoing picture does not necessarily hold when the experiments are repeated with sample 3 as we discuss next.  

%%%%%%%%%%%%
\begin{figure*}[t!]
    \centering
{\includegraphics[width=0.4\textwidth]{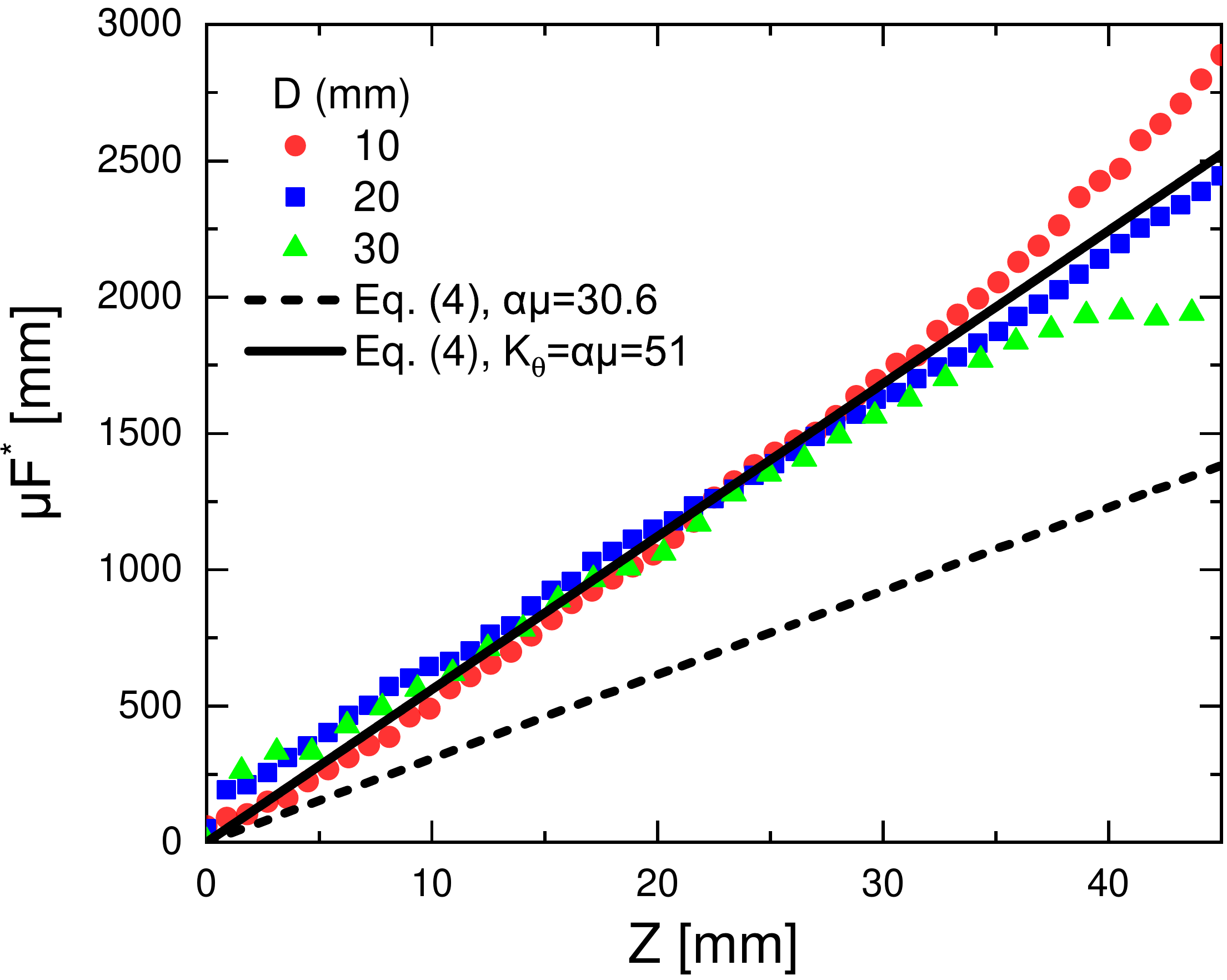}\label{panelA100}} (a) \hskip10mm      {\includegraphics[width=0.4\textwidth]{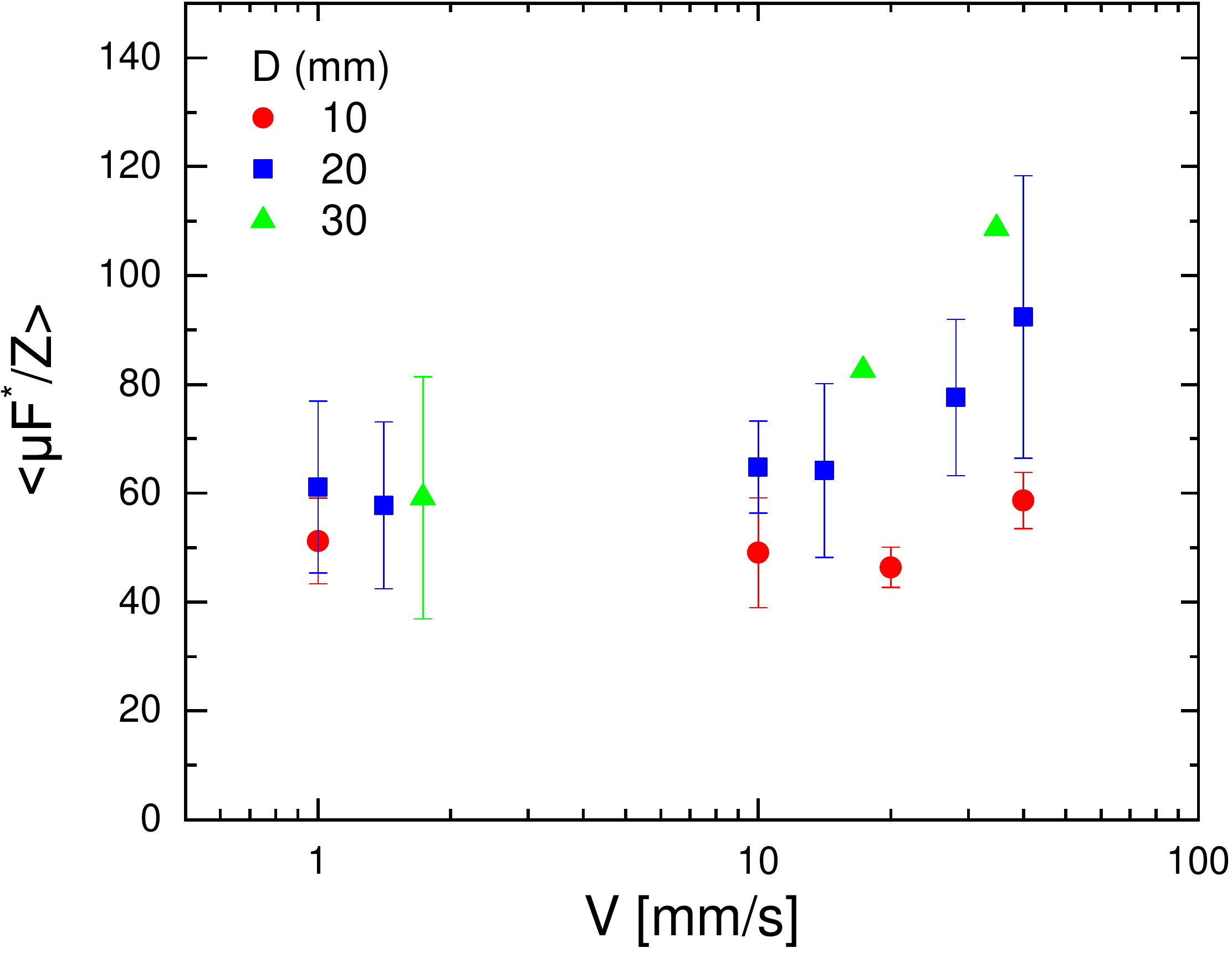}\label{panelB100}} (b) 
    \caption{(a) Rescaled drag force $\mu F^*=F/(\rho gR^2)$ as a function of the intrusion depth for  cylinders of different diameters axially penetrating (Fig. \ref{fig:schematic}b) granular beds of sample 3 at constant speed $V=1$ mm/s. Cylinder diameters are indicated on the graph. The dashed and solid lines show best fits to model predictions from Refs. \cite{brzinski2013depth,kang2018archimedes} as detailed in the text. (b) Semilogarithmic plot for  $\langle \mu F^*/Z\rangle$ as a function of the intrusion speed.}
    \label{100}
\end{figure*}
%%%%%%%%%%%%

%%%%%%%%%%%%
\begin{figure*}
    \centering
{\includegraphics[width=0.4\textwidth]{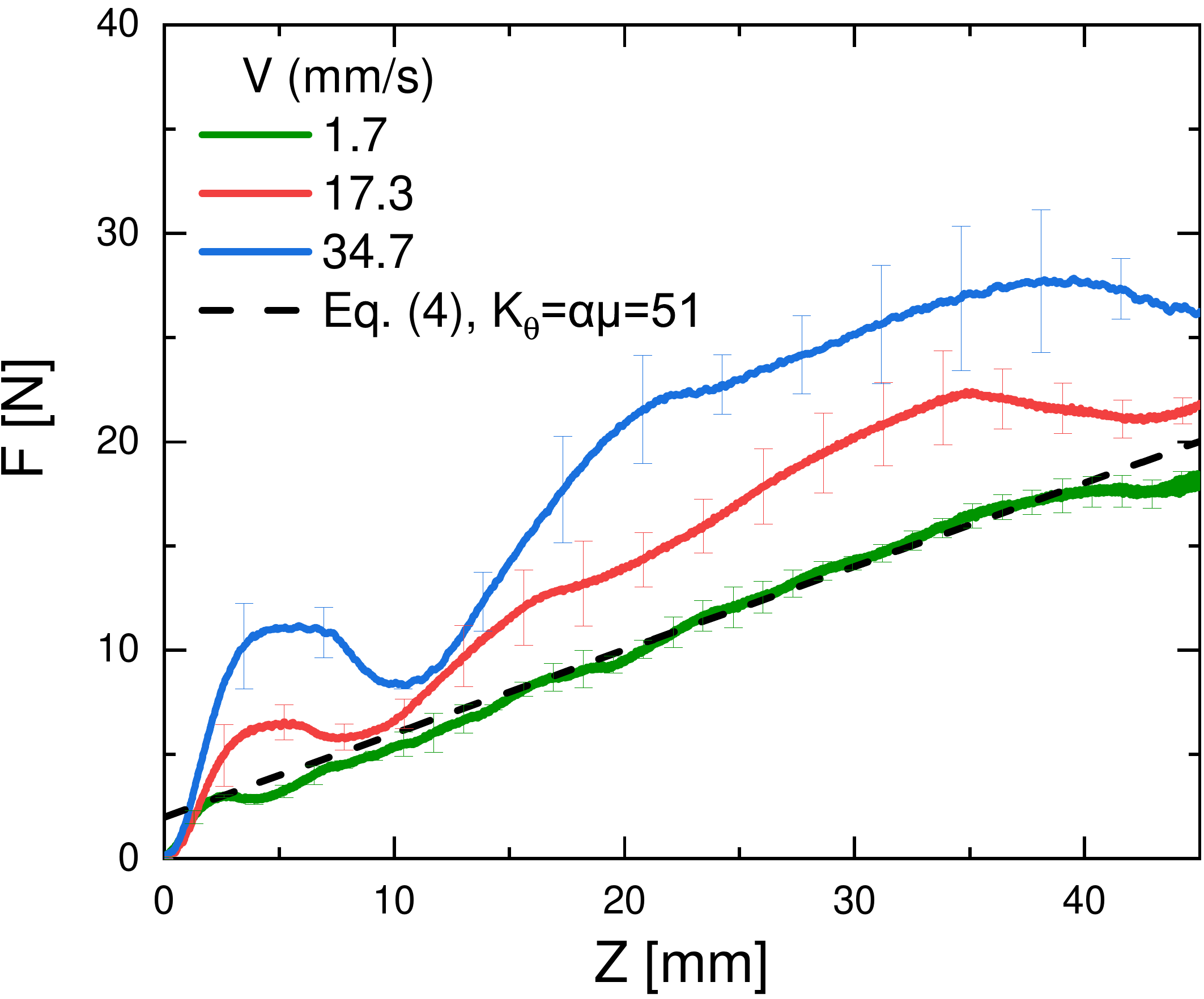}\label{d2}} (a) \hskip10mm
%  \hfill
{\includegraphics[width=0.4\textwidth]{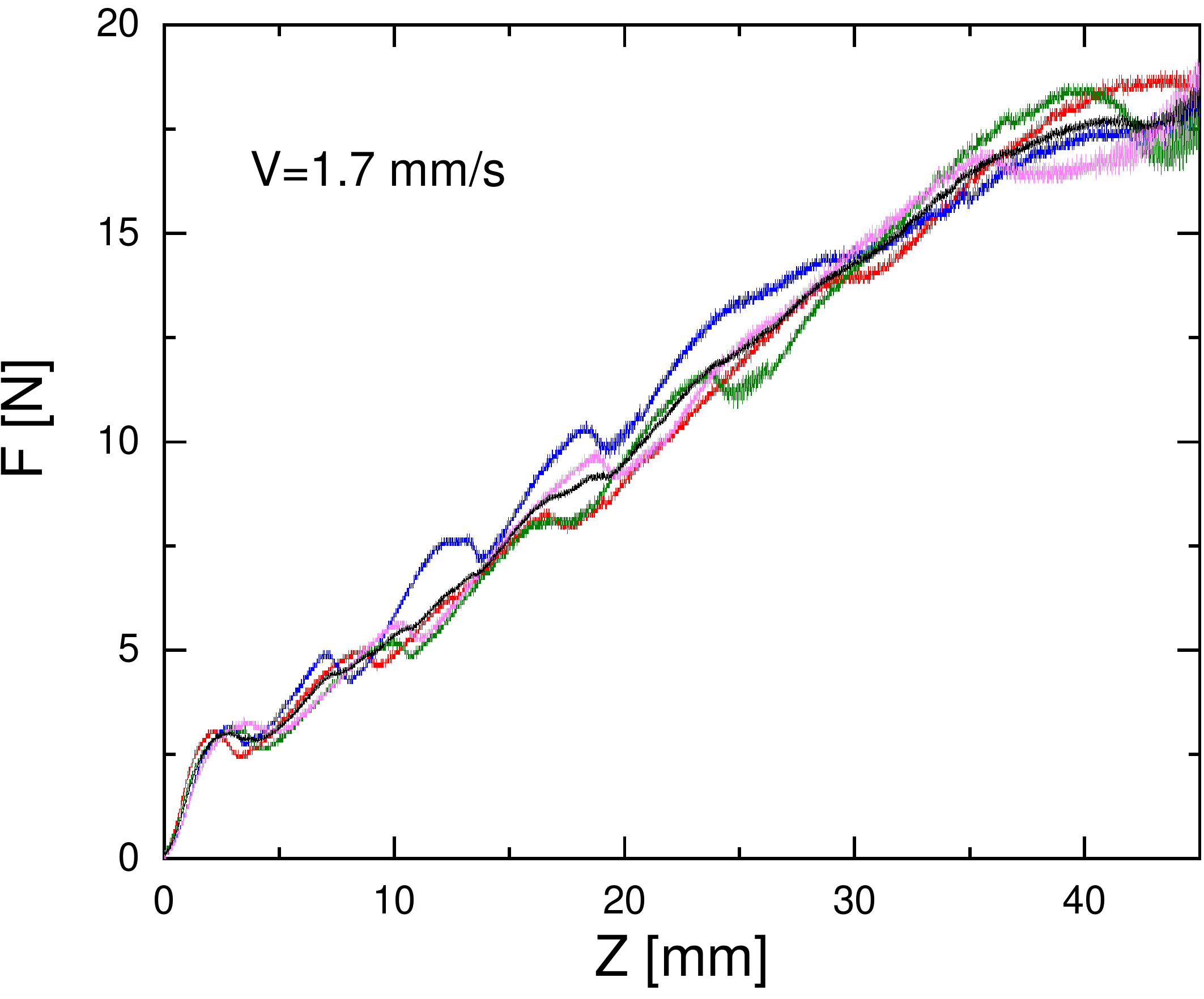}\label{d3}} (b) 
%  \hfill  
    \caption{(a) Drag force as a function of depth for different intrusion speeds  $V=1.7, 17.3$, and 34.7~mm/s for a cylinder of fixed diameter of $D=30$~mm axially penetrating granular beds of sample 3. (b) Non-averaged drag force from five individual experimental runs which average out to  the green curve in panel (a).}
    \label{Fig::MeanVal}
\end{figure*}
%%%%%%%%%%%%

%%%%%%%%%%%%%%%%%%%%%%%%%%%%%%%%%%%%
\subsection{High polydispersity effects and force-depth undulations}

Figure \ref{panelA100} shows the measured drag for sample 3 with the same parameter values as in Fig.  \ref{PA2,400} except that we now plot the quantity $\mu F^\ast=F/(\rho gR^2)$ as a function of the intrusion depth $Z$ (the redefinition is done to enable quantitative comparisons with the theoretical models of both Refs. \cite{brzinski2013depth,kang2018archimedes} in  one graph). As seen from the figure, the data show higher rate of increase for the drag force than  predicted by Ref. \cite{brzinski2013depth},  dashed line, but they are in close agreement with the modified model of Ref.  \cite{kang2018archimedes}, solid line. For the dashed line, we have used $\alpha\mu=30.6$ (based on $\alpha=60$ from Fig. \ref{PA2,400} and $\mu =0.51$  for sample 3) and, for the solid line, we have  extracted the factor $\alpha\mu=K_\theta\simeq 51$  directly from Fig. 5 of Ref. \cite{kang2018archimedes}. These values are then used in Eq. \eqref{VERTICAL} to depict the theoretical lines. 

The correction proposed in Ref. \cite{kang2018archimedes} clearly applies to our data from sample 3 and this is consistent with the angle of repose $\theta=27^{\circ}$ in this particular case. It is important to note that the data in Fig. \ref{panelA100} and, hence, the aforesaid agreement, are obtained for the  low intrusion speed of 1 mm/s. Figure \ref{panelB100} indeed reveals that, except for the smallest cylinder diameter ($D=10$~mm), the validity of quasistatic model breaks down in this sample when the speed is increased above 10~mm/s  (compare Figs. \ref{PB2,400vel} and \ref{panelB100}). Specifically, for cylinder diameters $D=20$ and 30~mm,   the mean drag force can increase by 50\% when the  speed is increased from 1 to 40 mm/s. This necessitates a closer examination of the velocity-dependence effects.

Figure \ref{d2} shows the drag force for the cylinder diameter $D=30$~mm for three different intrusion speeds $V=1.7, 17.3$, and 34.7 mm/s. The data for the smallest intrusion speed (green curve) maintain the same level of agreement with the corrected theoretical model of Ref. \cite{kang2018archimedes}  (dashed black line) as noted above. There appears weak undulations in the measured data that manifest themselves more clearly when the cases with higher intrusion speeds are examined (red and blue curves). The experimental curves and the error bars here are acquired from averaging over 5 repetitions as noted before. To unmask the nature of the underlying undulations in force-depth profiles  and avoid artifacts from the data averaging, we turn to the  profiles obtained directly from individual runs. These are shown in Fig. \ref{d3} for the low-speed set with $V=1.7$~mm/s. A clear pattern of alternating local peaks hence emerge. Such patterns are reproduced in other intrusion speeds as well (not shown). They indicate that the force is amplified and reaches a local maximum before it drops off relatively rapidly. This occurs  in a way that the whole curve maintains a nearly linear increase with the intrusion depth. 

The aforementioned trait in the force-depth profiles is reminiscent of general shear failure in soil mechanics  \cite{Terzaghi1943physics} and the local peaks resemble the ultimate bearing capacity, $F_\text{ult}$. The latter refers to the specific ability of the sample against tension before failure. In other words, as the vertical cylinder penetrates the granular bed at constant speed, the drag force increases to the critical value $F_\text{ult}$ at specific depths where shear failure surfaces form in the sample, extending away from bottom of the cylinder up to the granular surface. We indeed observe such failure events in the course of intrusion in sample 3. These are accompanied by the rapid reductions in the measured force as seen in Fig. \ref{d3}.

%%%%%%%%%%%%
\begin{figure}[t!]
    \centering
    \includegraphics[width=0.4\textwidth]{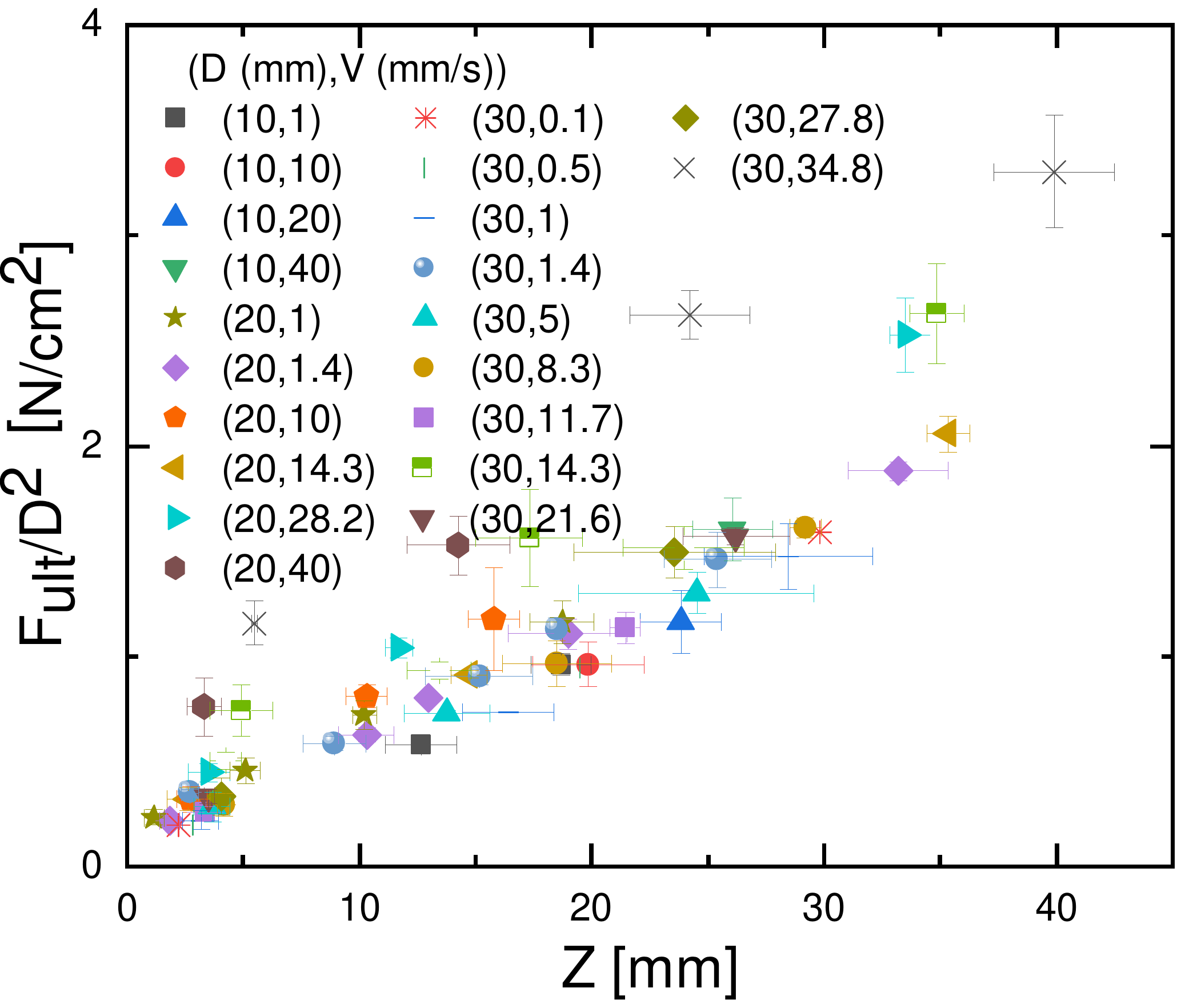}
    \caption{Rescaled peak force $F_\text{ult}/D^2$ and the corresponding depth, where the peak arises upon undulations of the force-depth profile in sample 3, are extracted and shown here for the whole set of cylinder diameters ($D$) and intrusion speeds ($V$) considered in our experiments.}
    \label{Fultra2}
\end{figure}
%%%%%%%%%%%%

%%%%%%%%%%%%
\begin{figure}
    \centering
{\includegraphics[width=0.4\textwidth]{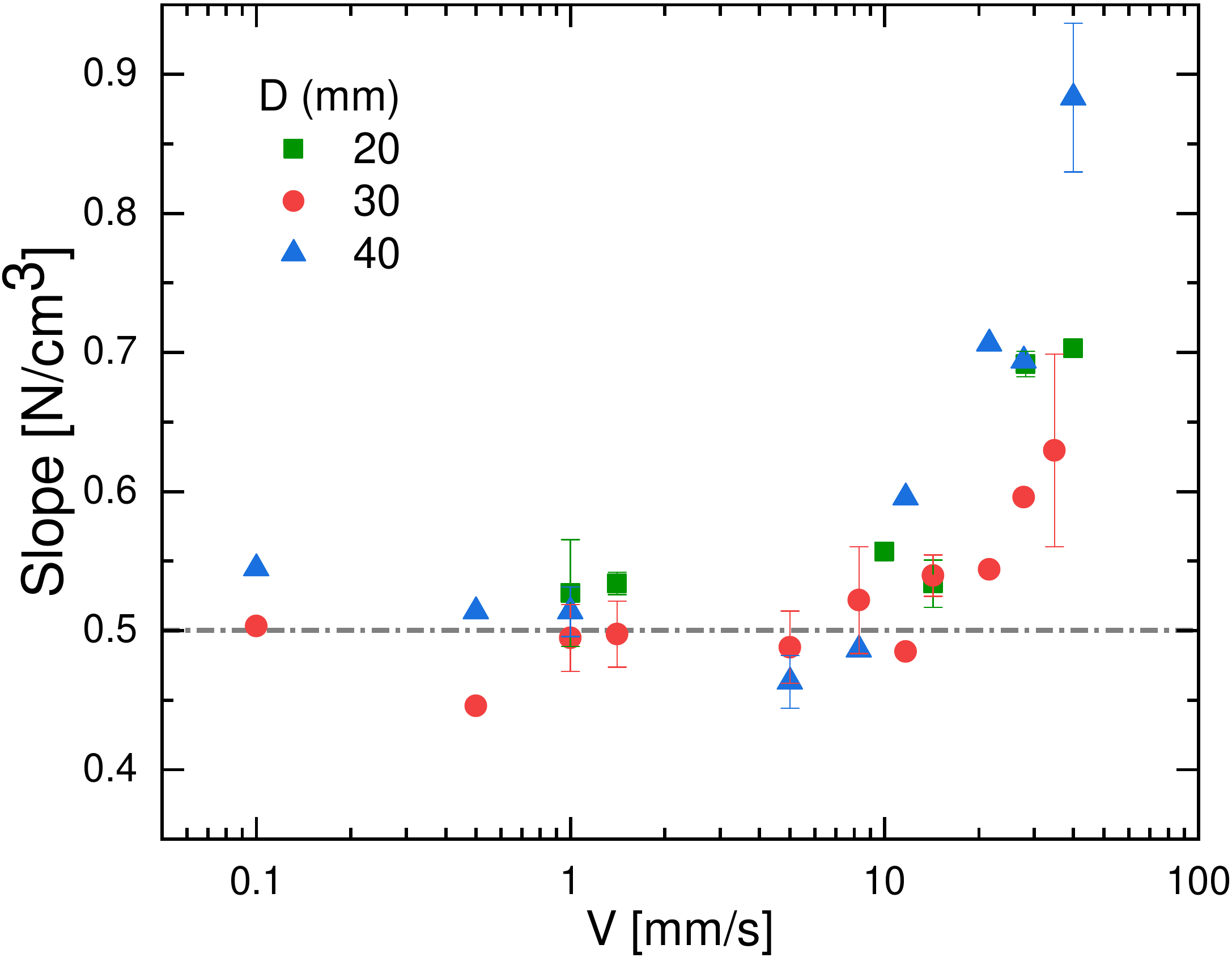}\label{slope}} (a) 
\hskip10mm
{\includegraphics[width=0.4\textwidth]{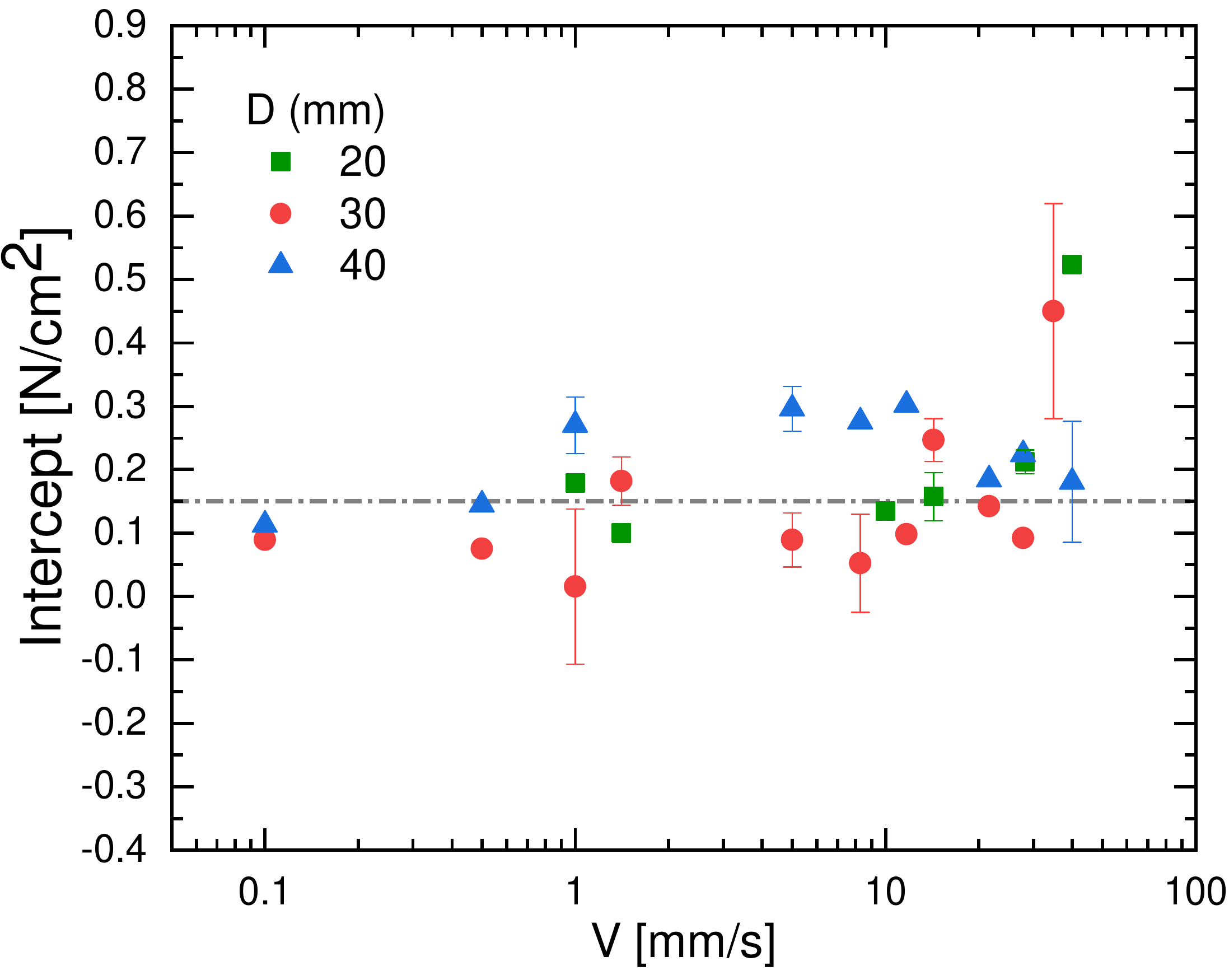}\label{intercept}} (b) 
    \caption{Semilogarithmic plots for (a) slope and (b) intercept of linear best fits to each data set in Fig. \ref{Fultra2} as functions of the intrusion speed.}
    \label{Fig::Slopes}
\end{figure}
%%%%%%%%%%%%

 According to the standard picture  \cite{Terzaghi1943physics}, for a shallow circular foundation whose depth is smaller than four times its diameter (as applicable to our experiments), the ultimate bearing force can be modeled as 
\begin{equation}
F_\text{ult}=\big(1.3\,CN_\text{c}+\gamma N_\text{q}Z+0.3\gamma DN_\gamma\big)\frac{\pi D^2}{4}
\label{Terzaghi}
\end{equation}
where $N_\text{c}$, $N_\text{q}$, and $N_\gamma$ are dimensionless bearing capacity factors and $C$  and $\gamma$ are the effective cohesion and specific weight of the sample, respectively. Because the intruder diameter in the present context is relatively small, the third term in Eq. \eqref{Terzaghi} turns out to be negligible in comparison with the first term. When the bearing capacity is rescaled as $F_\text{ult}/D^2$,  Eq.  \eqref{Terzaghi} predicts that the rescaled quantity displays a linear increase with the intrusion depth $Z$. More importantly, the slope and intercept of this linear behavior are independent of both the cylinder diameter and the  intrusion speed. The latter in fact mirrors the quasistatic condition under which the failure process is studied. Hence, in accord with our earlier findings (Fig. \ref{panelB100}), the linear form set forth by Eq.  \eqref{Terzaghi} could emerge here only for intrusion speeds up to 10~mm/s. 

To proceed, we extract the measured force at the local peaks of the force profile, $F_\text{ult}$,  and the depths where each peak arises. In Fig. \ref{Fultra2}, we show these data for the whole set of intrusion speeds and three different cylinder diameters that we have explored in the experiments on sample 3. The rescaling of the peak force data as $F_\text{ult}/D^2$ indeed corroborates the linear behavior predicted by Eq.  \eqref{Terzaghi}. To quantify the corresponding slope and intercept (while bearing in mind that the third term is small compared to the first one), we perform a linear best fit for each data set displayed in Fig. \ref{Fultra2}. Figures \ref{slope} and \ref{intercept} show the  slopes and intercepts of fitting lines as functions of the intrusion speed, respectively. As seen, both  these parameters remain roughly constant for intrusion speeds up to 10~mm/s and for the three shown cylinder diameters. This lends support to the proposed scenario that the force undulations result from general shear failures in the sample.  The agreement however remains semiquantitative. To clarify this point, we note that our data in Figs. \ref{slope} and \ref{intercept}  indicate line slope and intercept of around  0.5~$\text{N/cm}^3$ and 0.15 $\text{N/cm}^2$ (dot-dashed lines) within the aforementioned speed range. On the other hand, based on Mohr-Coulomb failure criterion \cite{MICHALOWSKI199757}, direct shear tests yield the cohesion of sample 3 as  $C\simeq 0.1656\,\text{N/cm}^2$. Also, assuming $\theta=27^{\circ}$, $N_\text{c}$, $N_\text{q}$, and $N_\gamma$ can be read off from tabulated values in Ref. \cite{Terzaghi1943physics} as 30, 16 and 12, respectively. These give the slope and intercept predicted from the bearing capacity model, Eq. \eqref{Terzaghi},  as  $\pi\gamma N_\text{q}/4\simeq 0.16\,\text{N/cm}^3$ and $1.3\pi CN_\text{c}/4\simeq 5.07\,\text{N/cm}^2$. These  values differ from the measured slope and intercept in our experiments as  noted above. Despite the overall agreement between our findings and Eq. \eqref{Terzaghi}, such quantitative differences between the data and the model considered here remain to be clarified. It is worth noting that, since the original work of Ref. \cite{Terzaghi1943physics}, different approaches have been attempted at modeling the bearing capacity and the issue remains subject to ongoing scrutiny; see, e.g., Refs. \cite{MOTRA2016265,PENG2019601,ZHANG2020103734,Chen2020,ahmad2021prediction} and references therein. Numerical simulations will be of value to illuminate the underlying  mechanisms at work in the present system but they go beyond the scope of the current investigation. 

%%%%%%%%%%%%%%%%%%%%%%%%%%%%%%%%%%%%
%%%%%%%%%%%%%%%%%%%%%%%%%%%%%%%%%%%%
\section{Conclusion}

We investigated the drag force exerted on a cylindrical object intruding a granular bed of glass beads through a custom-made setup that allows for force measurement at constant penetration speed. In a horizontal quasi-2D setting, where the cylinder laterally penetrates the granular bed, we explored the regimes of hydrostatic-like behavior, force saturation and exponential growths (at shallow, intermediate and large depths close to the container bottom, respectively). We made  quantitative comparisons with theoretical predictions that, to our knowledge, had remained unaddressed in the literature. By focusing on the shallow-depth regime, we studied the effects of high grain polydispersity on the drag force also in a case where the cylinder axially penetrates the granular bed. In this case, we unveiled distinct qualitative differences in the force-depth profiles between the drag force in a highly polydisperse sample and two other relatively monodisperse samples. In the highly polydisperse sample, our data for the averaged drag force in the quasistatic regime agree well with the modified frictional drag model proposed in Ref. \cite{kang2018archimedes}, which is obtained as an extension of the original work by Brzinski  et. al \cite{brzinski2013depth}. Our results also reveal pronounced undulations in the force-depth profiles, which we argue to arise from consecutive general shear failure events that take place in the medium as the intrusion proceeds further. This leads to peak forces, representing the ultimate bearing capacity of the sample, at specific depths which we analyzed  in detail for a wide range of intrusion speeds and for different cylinder diameters. We showed that the peak forces scale quadratically with the cylinder diameter. For a given cylinder diameter, the peak forces grow linearly with depth. The slope and intercept of this linear behavior are independent of the intrusion speed up to values around 10~mm/s. When rescaled with the squared diameter (cylinder base area), the slope and intercept appear to be independent of the diameter. These aspects agree with the ultimate bearing capacity theory in soil mechanics even as  model predictions for the  slope and intercept exhibit quantitative deviations from our experiments. Such deviations warrant further investigations into the mechanisms at work within the present context. Our data in the case of sample 3 also indicate  deviations from the quasistatic picture at intrusion speeds above 10~mm/s. It is important to note that, due to a broad distribution of grain sizes in the sample (1-100~$\upmu$m), inertial contributions may come into play. This is because the critical velocity itself admits a distribution of values laying below and above the intrusion velocity (for sufficiently small and large grains, respectively). This necessitates an in-depth theoretical and/or numerical modeling of the problem when the sample includes a wide range of grain sizes. The latter presents an interesting direction for future research given that high polydispersity effects are expected to be relevant in numerous areas such as geotechnical applications, subsurface robotic excavation, industrial drilling, etc. Other areas of interest that can be explored in the context of highly polydisperse samples include effective interactions between rotating objects and surrounding grains in mixing and segregation applications \cite{ZHAO20201,LIU201741} as well as presence of interstitial liquids which can directly affect material resistance \cite{M_ller_2007,pakpour2012construct,RADL2010171,ARTONI2019231}. 

\section{Author contributions}
{\bf Salar Abbasi Aghda:} Conceptualization, Methodology, Validation, Formal analysis, Investigation, Writing - Original draft, Visualization. {\bf Ali Naji:} Conceptualization, Methodology,  Writing - Original draft, Supervision, Project administration. 

\section{Acknowledgements}
We acknowledge useful communications with T. A. Brzinski and A. Seguin and also discussions with S. M. Taheri and M. Pakpour at initial stages of this work. This research did not receive any specific grant from funding agencies in the public, commercial, or not-for-profit sectors. 

\section{Conflicts of interest}
The authors declare that they have no known competing financial interests or personal relationships that could have appeared to influence the work reported in this paper.
\bibliography{bibliography}

\end{document}